\documentclass[global]{svjour}

\usepackage{graphics}

\usepackage{colortbl}

\journalname{myjournal}
\begin{document}
\title{Blue laser cooling transitions in Tm I}

\author{ N.~Kolachevsky \and A.~Akimov \and I.~Tolstikhina \and K.~Chebakov \and A.~Sokolov \and P.~Rodionov \and S.~Kanorski \and V.~Sorokin
}
\institute{P.N. Lebedev Physics Institute, Leninsky prospekt 53,
119991 Moscow, Russian Federation\\ E-mail: kolik@lebedev.ru, Fax:
+7 495 1326644 \\ Pacs: 32.70.Cs, 32.10.Fn, 32.80.Pj}
\date{Received: date / Revised version: date}
\maketitle
\begin{abstract}
We have studied possible candidates for laser cooling transitions
in $^{169}$Tm in the spectral region 410\,--\,420\,nm. By means of
saturation absorption spectroscopy we have measured the hyperfine
structure and rates of two nearly closed cycling transitions from
the ground state
$4\textrm{f}^{13}6\textrm{s}^2(^2\textrm{F}_0)(J_g=7/2)$ to upper
states
$4\textrm{f}^{12}(^3\textrm{H}_5)5\textrm{d}_{3/2}6\textrm{s}^2(J_e=9/2)$
at $410.6$\,nm and
$4\textrm{f}^{12}(^3\textrm{F}_4)5\textrm{d}_{5/2}6\textrm{s}^2(J_e=9/2)$
at 420.4\,nm and evaluated the life times of the excited levels as
{15.9(8)\,ns} and {48(6)\,ns} respectively. Decay rates from these
 levels to neighboring opposite-parity levels are evaluated
by means of Hartree-Fock calculations. We conclude, that the
strong transition at $410.6$\,nm has an optical leak rate of less
then $2\cdot10^{-5}$ and can be used for efficient laser cooling
of $^{169}$Tm from a thermal atomic beam. {The hyperfine structure
of two other even-parity levels which can be excited from the
ground state at 409.5\,nm and 418.9\,nm is also measured by the
same technique.} In addition we give a calculated value of
$7(2)$\,s$^{-1}$ for the rate of magnetic-dipole transition at
1.14\,$\mu$m between the fine structure levels
$(J_g=7/2)\leftrightarrow(J'_g=5/2)$ of the ground state { which
can be considered as a candidate for applications in atomic
clocks}.
\end{abstract}
\section{Introduction}
During the last decade, significant progress has been achieved in
laser cooling of lanthanides (rare-earth elements). Laser-cooled
lanthanides are effectively used in such fundamental fields as the
study of cold collisions \cite{Santos},  Bose-Einstein
condensation \cite{Takasu}, ultra-precise atomic clocks
\cite{Taichenachev} and also open new perspectives for
implementation in nano-technology \cite{Hill} and quantum
information \cite{Monroe}. In contrast to recently demonstrated
method of buffer gas cooling and trapping of lanthanides in a
magnetic dipole trap \cite{Hancox}, laser-cooled atoms are easily
manipulated by the help of light fields \cite{Katori} and can be
studied in a nearly perturbation-free regime.

Compared to atoms from the alkali and the alkali-earth groups,
 spectra of lanthanides are significantly richer due to the presence of
the 4f shell electrons. Ytterbium with its closed 4f$^{14}$ shell
possesses the simplest level structure and has been successfully
laser cooled at the wavelength of 398.9\,nm (see e.g.
\cite{Maruyama}). In 2006 cooling of atomic erbium was reported
\cite{McClelland} at the wavelength of 401\,nm. Both these strong
cooling transitions are not completely closed and their upper
levels also decay to the neighboring opposite-parity levels thus
taking  a part of population out from the cooling cycle (optical
leaks). In the case of erbium (4f$^{12}$6s$^2$), evaluation of a
leak rate is a complex task due to a rich level structure. Still,
it has been shown experimentally, that it is possible to cool and
to trap up to $10^6$ erbium atoms  in a magneto-optical trap (MOT)
even without a repumper laser.

In this paper, we analyze the possibility to cool atomic thulium
which resides between Er and Yb in the periodic table. Since there
is only one unfilled electron in the 4f shell (the ground state of
Tm has a configuration of
$4\textrm{f}^{13}6\textrm{s}^2(^2\textrm{F}_0)$), its electronic
structure is more complex than that of Yb, but still is one of the
simplest among lanthanides. Thulium has only one stable isotope
$^{169}$Tm with a nuclear spin number of $I=1/2$ which results in
a simple doublet hyperfine splitting of each electronic level. The
monoisotopic structure should increase a MOT loading rate, { while
the non-degenerate Zeeman structure of the ground level
 enables sub-Doppler cooling
schemes}.

The ground state of $^{169}$Tm consists of two fine structure
levels with total electronic momentum numbers of $J_g=7/2$ and
$J'_g=5/2$ which are separated by
 $2.6\cdot10^{14}$\,Hz (the corresponding transition wavelength
 is
 $\lambda=1.14\,\mu$m). The excited ground-state level with
 $J'_g=5/2$ is metastable and one can expect its life time to be on the
 order of a few fractions of a second \cite{Ban}.
{Due to the shielding by the  outer closed  6s$^2$ shell
\cite{Hancox,Aleksandrov}, it is expected that interrogation of
these forbidden
 transitions even in dense laser-cooled atomic clouds will allow
 one
 to build precise optical atomic references possessing a high
 short-term stability.}

 In this paper we describe our
experimental study of two {candidates} for cooling transitions
from the ground state to the excited states
$4\textrm{f}^{12}(^3\textrm{H}_5)5\textrm{d}_{3/2}6\textrm{s}^2(J_e=9/2)$
at $410.6$\,nm and
$4\textrm{f}^{12}(^3\textrm{F}_4)5\textrm{d}_{5/2}6\textrm{s}^2(J_e=9/2)$
at 420.4\,nm (Section~II). {The hyperfine structure (HFS) of these
levels are accurately determined from the experiment.}
{Simultaneously we have for the first time determined the HFS of
two other excited levels which are within reach of our laser
system.} From the analysis of saturation absorption spectra we
have experimentally determined lifetimes of the excited levels and
compared them with existing data \cite{Wickliffe}. With the
relativistic Hartree-Fock code of Cowan \cite{Cowan}, optical leak
rates are quantitative evaluated. Results of this analysis are
presented in Section~III. In Section~IV we discuss realistic laser
cooling schemes for thulium. In the last Section we analyze the
magnetic dipole transition at 1.14\,$\mu$m with the help of the
Cowan code and also with the flexible atomic code (FAC) of Feng
\cite{FAC}.

\section{Saturation absorption spectroscopy}

To efficiently laser cool an atomic sample one has to choose a
strong closed cycling transition which can be  excited by a
powerful laser source. In the case of thulium, we restrict our
consideration to transitions from the ground state level
$4\textrm{f}^{12}6\textrm{s}^2 (J_g=7/2)$ to the excited levels of
the opposite parity with $J_e=9/2$. Such a choice allows the use
of cycling transitions between the hyperfine components of the
lower and upper levels $F_g=4\leftrightarrow F_e=5$. In this case,
electric dipole transitions from upper levels to other fine and
hyperfine sublevels of the ground state will be forbidden by the
selection rules.

\begin{figure}[h!]
\resizebox{1 \textwidth}{!}{%
  \includegraphics{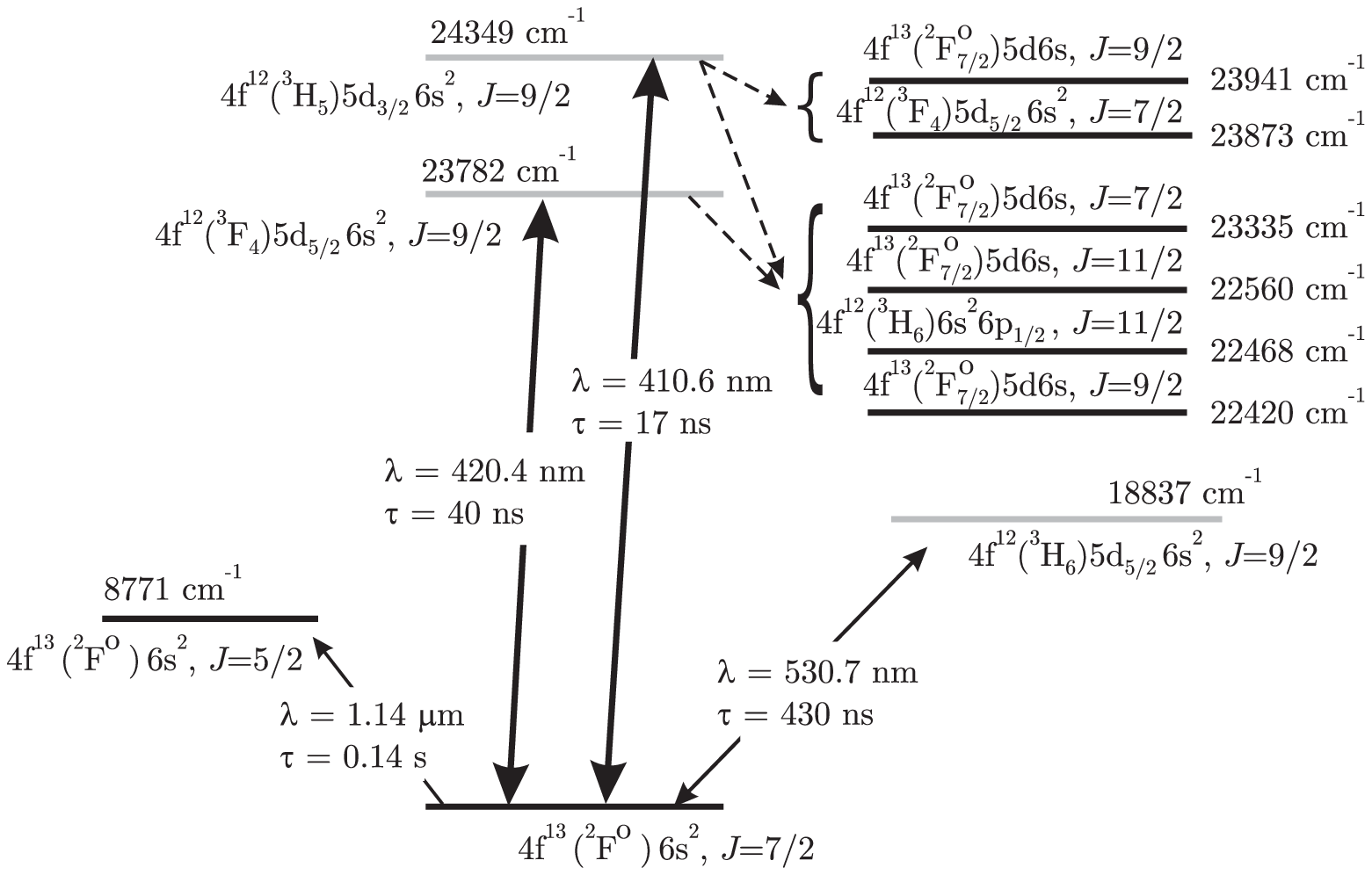}}
\caption{A partial level diagram of atomic thulium. Even and odd
levels are shown in { gray and black} respectively. Some
transition wavelengthes and the life times of the corresponding
excited levels are presented. }\label{fig1}
\end{figure}

Among excited levels in Tm one can select three candidates
favorable for laser cooling \cite{NISTdatabase}:
$4\textrm{f}^{12}(^3\textrm{H}_6)5\textrm{d}_{5/2}6\textrm{s}^2$
($E=18\,837\,\textrm{cm}^{-1}$),
$4\textrm{f}^{12}(^3\textrm{F}_4)5\textrm{d}_{5/2}6\textrm{s}^2$
($E=23\,782\,\textrm{cm}^{-1}$) and
$4\textrm{f}^{12}(^3\textrm{H}_5)5\textrm{d}_{3/2}6\textrm{s}^2$
($E=24\,349\,\textrm{cm}^{-1}$), which are schematically presented
in Fig.\,\ref{fig1}. The green transition at $\lambda=530.7$\,nm
can be excited by the second harmonic of a laser on gadolinium
scandium gallium garnet activated by neodymium (Nd:GSGG)
\cite{Zharikov}. The transition is completely closed in the
electric-dipole approximation which removes the problem of optical
leaks. Unfortunately, the low rate of
$A=2.3\cdot10^6\,\textrm{s}^{-1}$ prevents using this transition
for efficient loading a MOT from a thermal beam (an atom with
initial velocity of 200~m/s can be decelerated at a distance not
shorter then 3~m). Nevertheless it can be used after cooling an
atomic sample at other strong transition. For example, such a
 blue and green MOT sequence is implemented for cooling
ytterbium to sub-mK temperatures \cite{Maruyama}.

 Other transitions at 420.4\,nm and 410.6\,nm can be excited by the second harmonic
 of a titanium:sapphire (Ti:Sa) laser, the second harmonic of an infrared  diode laser
 or directly by a nitride diode laser (see e.g.
 \cite{Hult}). According to \cite{Wickliffe} these transitions have  rates of
$2.43\cdot10^7\,\textrm{s}^{-1}$  and
$6.36\cdot10^7\,\textrm{s}^{-1}$ respectively which should be
sufficient to decelerate hot atoms at a distance of a few tens of
centimeters. The transitions are not completely closed. The upper
levels are coupled to neighboring opposite-parity levels as shown
in Fig.\ref{fig1}. To choose the best candidate for cooling
transition it would be desirable to evaluate optical leak rates.

\begin{figure}[h!]
\resizebox{0.75 \textwidth}{!}{%
  \includegraphics{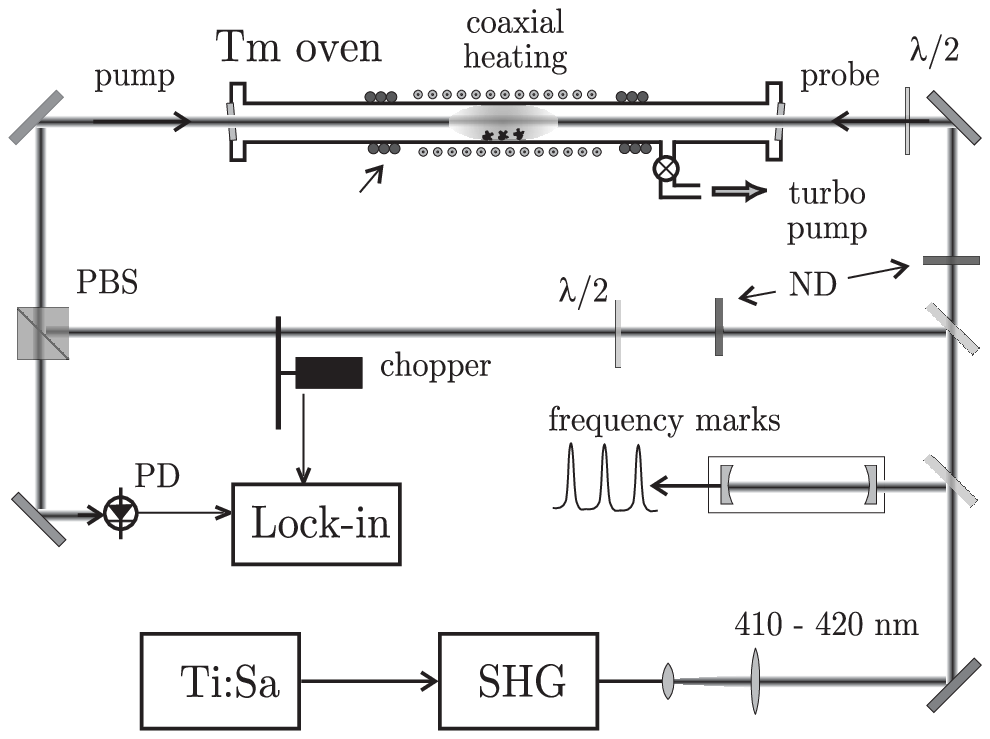}}
\caption{Saturation absorption spectroscopy setup with
lin$\perp$lin polarizations. Thulium is heated in a stainless
steel vacuum cell to 1000\,K by a coaxial heating cable. A stable
 interferometer is used to calibrate recorded spectra. Here
SHG is a second harmonic generation stage, PBS -- a polarization
beam splitter, PD -- a silicon photodetector, and ND -- variable
neutral density filters.}\label{fig2}
\end{figure}

{ The hyperfine structure of excited levels in thulium is measured
by different methods and the following references summarize most
of the available data \cite{Childs,Leeuwen,Pfeufer,Kroeger,Basar}.
The level with the energy $E=24\,349\,\textrm{cm}^{-1}$ has been
previously studied by the optogalvanic spectroscopy in the hollow
cathode discharge \cite{Kroeger} and its HFS frequency has been
determined. We use saturation absorption spectroscopy in
counter-propagating beams of the same frequency to measure the HFS
of the levels which can be excited from the ground state in the
wavelength region 410\,--\,420\,nm. }
 The experimental setup is
presented in Fig.\,\ref{fig2}.

Radiation of a Ti:Sa laser (MBR-110, Coherent Inc.) is frequency
doubled in a lithium triobate crystal (LBO) placed in an external
cavity (MBD-200, Coherent Inc.). The laser system produces up to
150\,mW of radiation in the spectral region 400\,--\,430\,nm. The
spectral line width of the Ti:Sa laser is specified as 100\,kHz
which is achieved by locking the laser to a stable high-finesse
Fabry-Perot cavity. We tune the laser to the atomic transition
with the help of a home-made wavemeter. The astigmatic blue laser
beam is expanded to about $w_{0,x}\times
w_{0,y}\approx3\,\textrm{mm}\times6\,\textrm{mm}$ (at
 $1/e^2$ level) and is split into saturation and probe beams. The beams
are carefully aligned in the counter-propagating configuration
with the angle between them less than $5\cdot10^{-4}$. The probe
beam is modulated by a wheel chopper at 850\,Hz, and the signal
from the probe beam is recorded using the lock-in technique by a
computer. The beams have orthogonal linear polarizations
(lin$\perp$lin) and are separated on a polarization beam splitter.
Intensities of the beams can be varied by absorptive neutral
density filters. They are measured by a calibrated power meter at
the cell entrance. The laser frequency detuning is controlled by a
stable confocal interferometer with a free spectral range of
75\,MHz. Frequency marks are recorded simultaneously with
absorption spectra.

Thulium vapor is produced in a stainless steel oven of 20\,mm in
diameter. The central part can be heated by a coaxial cable to
1100\,K. Current flowing through the central heating wire of the
cable returns back through its outer shielding which significantly
cancels out an induced magnetic field.  External coils in
Helmholtz configuration allow compensation of the laboratory
magnetic field to less then 1\,G. The melting point of thulium is
1818\,K, but a vapor pressure of 10$^{-3}$\,mbar can be reached
already at about 1000\,K. To prevent heating of the whole cell,
its central part is surrounded by water cooling coils. The cell is
pumped out by 20\,l/s turbo-molecular pump. Thulium chunks of a
few hundred milligrams are placed in the central part of the oven.
After heating to 1000\,K, we observe the absorption of 50\% in the
center of the Doppler-broadened line at 410.6\,nm. The pipe can
stably operate in this regime for days.

\begin{figure}[h!]
\resizebox{1 \textwidth}{!}{%
  \includegraphics{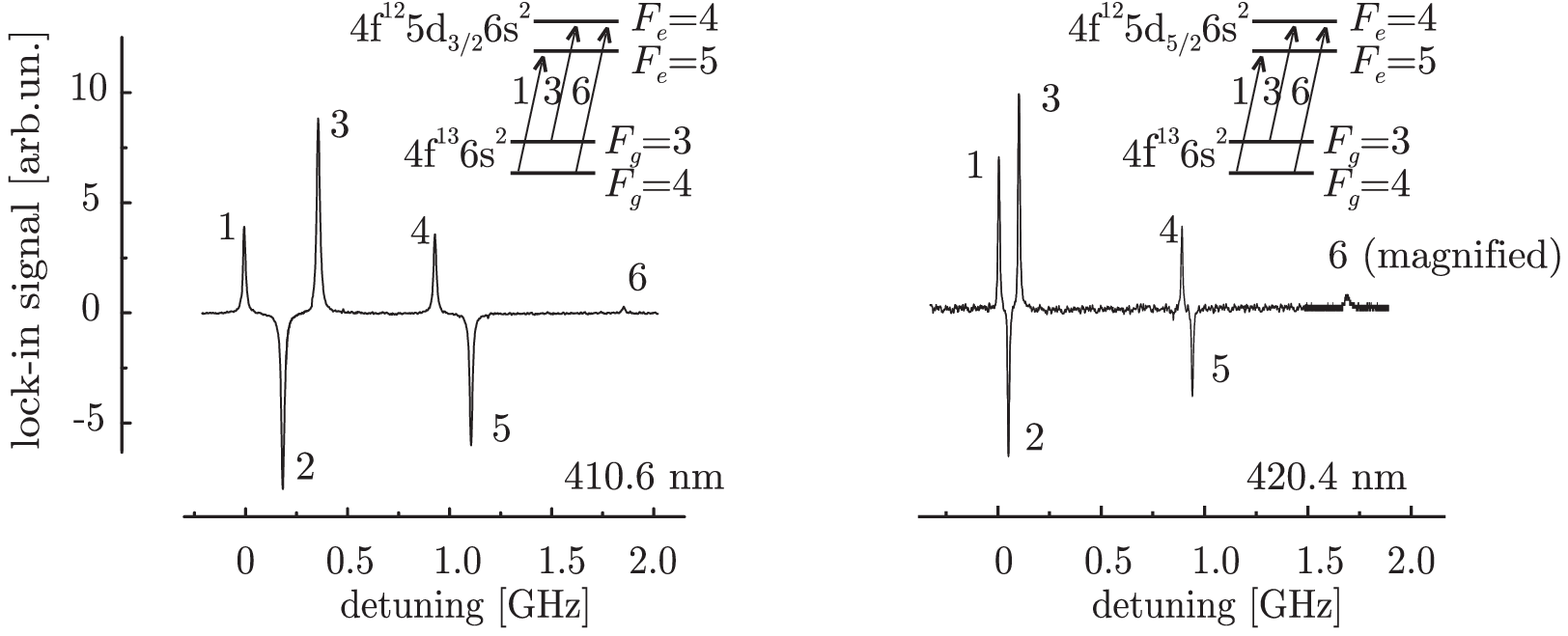}}
\caption{Saturation absorption spectra of the transitions in
$^{169}$Tm from the ground state
$4\textrm{f}^{13}6\textrm{s}^2(^2\textrm{F}_0)(J_g=7/2)$ to upper
states
$4\textrm{f}^{12}(^3\textrm{H}_5)5\textrm{d}_{3/2}6\textrm{s}^2(J_e=9/2)$
at $410.6$\,nm and
$4\textrm{f}^{12}(^3\textrm{F}_4)5\textrm{d}_{5/2}6\textrm{s}^2(J_e=9/2)$
at 420.4\,nm recorded in lin$\perp$lin polarizations. Along with
saturation absorption resonances, cross-over resonances are
observed. }\label{fig3}
\end{figure}

Figure \ref{fig3} presents saturation absorption spectra for the
transitions at 410.6\,nm and 420.4\,nm. Besides saturation
absorption lines formed by atoms with  zero velocity projection on
the beam axis, we observe cross-over resonances of different
signs. Identification of hyperfine spectra is  presented in the
insets of Fig.\,\ref{fig2}. The accurate result of
$-1496.550(1)$\,MHz for  the  HFS of the ground state
$4\textrm{f}^{12}6\textrm{s}^2\,(J_g=7/2)$ measured by Childs {\it
et al.} \cite{Childs} is used to calibrate the the free spectral
range of the Fabry-Perot cavity.  Here we will use negative
frequencies for the HFS if the Fermi energy is negative.

To determine the hyperfine structure frequency of the excited
levels we fit the recorded spectra by a multi-peak Lorentzian
function and use the frequency ruler of the Fabry-Perot cavity.
Uncertainty of the measurement is the sum of the statistical
uncertainty of 0.5\,MHz and the systematic contribution of
0.5\,MHz. The latter mainly results from the asymmetry of
trasmission peaks of the  {Fabry}-Perot cavity and is evaluated by
measuring positions of the cross-over resonances. No dependence of
the HFS frequencies on light intensity is observed. Results of the
measurement are presented in table \ref{tab1}.

It is interesting to note that the hyperfine splitting of upper
and lower levels of both candidates for cooling transitions  are
similar and  differ for only a few tens of natural line widths. It
can result in the situation, that a laser tuned to the red wing of
the the cooling transition~1 (Fig.\,\ref{fig3}) will
simultaneously transfer population from $F_g=3$ via the
transition~3. Due to the relatively small detuning, the process
will have a higher probability than the non-resonance population
transfer to the $F_g=3$ level via $F_e=4$ one. Laser cooling
without repumping has been successfully demonstrated in erbium
\cite{McClelland}.

\begin{table}[b!]
\caption{{The hyperfine splitting frequencies of four excited
levels in atomic thulium. The ground state HFS is also given for
the reference.}}\label{tab1}
\begin{tabular}{l@{\hspace{5ex}\ }l@{\hspace{5ex}\ }c@{\hspace{5ex}\ }l@{\hspace{5ex}\ }c}
\hline\noalign{\smallskip}
  Energy,  & Level   &  $J$ & HFS splitting,& Reference \\
 cm$^{-1}$ &  &  & MHz &\\

\noalign{\smallskip}\hline\noalign{\smallskip}
   0&$4\textrm{f}^{13}6\textrm{s}^2(^2\textrm{F}_0)$&$7/2$&$-1496.550(1)$&\cite{Childs}\\
23\,781.698& $4\textrm{f}^{12}(^3\textrm{F}_4)5\textrm{d}_{5/2}6\textrm{s}^2$ &9/2& $-1586.6(8)$&this work\\
 {23\,873.207} & ${4\textrm{f}^{12}(^3\textrm{F}_4)5\textrm{d}_{5/2}6\textrm{s}^2}$ & {7/2}&${+1411.0(7)}$& {this work}\\
  24\,348.692& $4\textrm{f}^{12}(^3\textrm{H}_5)5\textrm{d}_{3/2}6\textrm{s}^2$& 9/2& ${-1856.5(2.5)}$ &{ \cite{Kroeger}}\\
& & & $-1857.5(8)$ & this work\\
{24\,418.018}&
${4\textrm{f}^{13}(^2\textrm{F}^\circ_{7/2})6\textrm{s}6\textrm{p}(^1\textrm{P}^\circ_{1})}$
& {5/2} &${-1969.4(1.3)}$&{this work}\\\noalign{\smallskip}\hline
\end{tabular}
\end{table}

 The cooling transition rate is a crucial parameter which defines
the maximal achievable deceleration of atom as well as  the
Doppler cooling limit \cite{McClelland1}. We use the setup shown
in Fig.\,\ref{fig2} to measure the natural line width of
candidates for cooling transitions $F_g=4\leftrightarrow F_e=5$ at
410.6\,nm and 420.4\,nm (see insets in Fig.\ref{fig3}). Being
excited at these nearly closed cycling transitions, thulium
behaves as a two-level system which allows us to neglect optical
pumping and coherent effects  \cite{Letokhov}.

\begin{figure}[h!]
\resizebox{1\textwidth}{!}{%
  \includegraphics{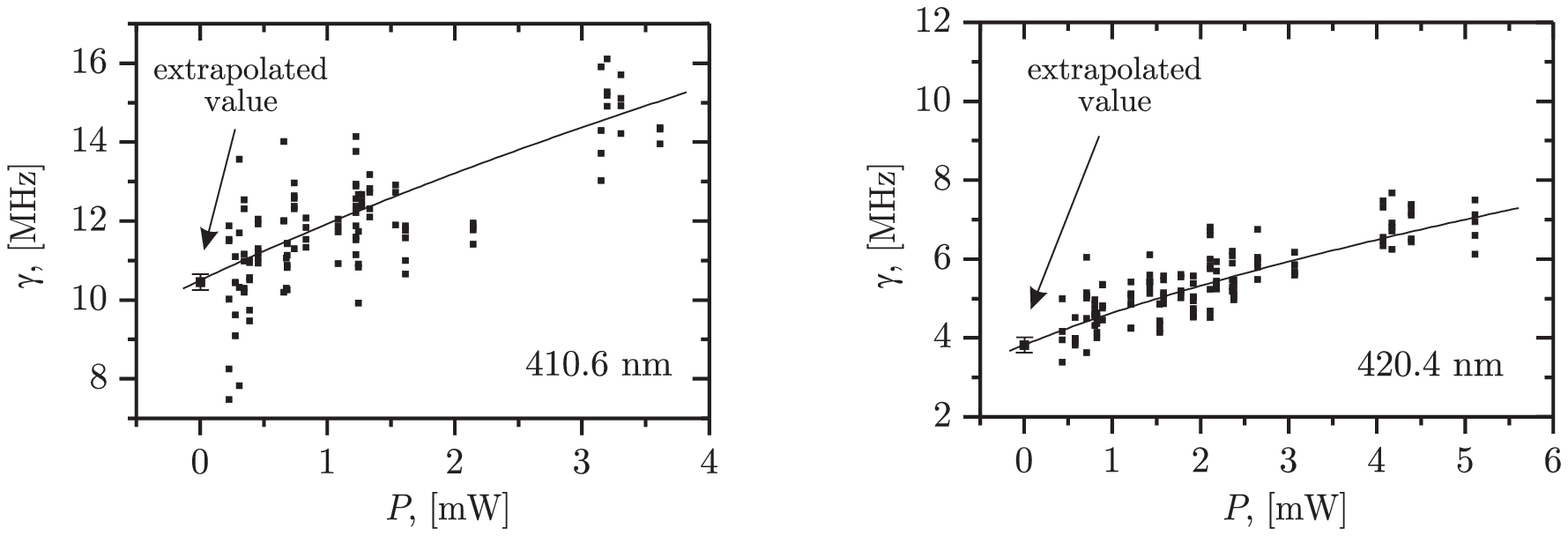}}
\caption{ Spectral line widths $\gamma$ of the transitions
$4\textrm{f}^{13}6\textrm{s}^2(^2\textrm{F}_0)(F=4)\leftrightarrow4\textrm{f}^{12}(^3\textrm{H}_5)5\textrm{d}_{3/2}6\textrm{s}^2
(F=5)$  (left, 410.6\,nm) and
$4\textrm{f}^{13}6\textrm{s}^2(^2\textrm{F}_0)(F=4)\leftrightarrow4\textrm{f}^{12}(^3\textrm{F}_4)5\textrm{d}_{5/2}6\textrm{s}^2(F=5)$
(right, 420.4\,nm) versus the excitation power $P$.
 }\label{fig4}
\end{figure}

We performed a set of measurements varying saturation and probe
power densities.  The power of 1\,mW approximately corresponds to
the on-axis power density of about 3\,mW/cm$^2$. Results are
presented in Fig.\,\ref{fig4}, where the measured line width
$\gamma$ is plotted against the excitation power $P$. Again, each
recorded spectrum has been fitted by a multi-peak Lorentzian
function with independent fit parameters for each peak. In the
case of a weak probe beam, the line width $\gamma$ is given by the
following expression \cite{Pappas,Ohshima}
\begin{equation}\label{eq1}
\gamma(I)=\frac{1}{2}\gamma_0(1+\sqrt{1+I/I_\textrm{sat}})\,,
\end{equation}
where $\gamma_0$ is the natural line width and $I$ is an
excitation power density. The saturation power density
$I_\textrm{sat}$ is defined as $I_\textrm{sat}=2\pi h c
\gamma_0/3\lambda^3$. In the general case the expression is more
complex \cite{Letokhov}, but the relation $\gamma(0)=\gamma_0$
remains valid.

We fit the data presented in Fig.\,\ref{fig4} by the function
(\ref{eq1}), where $\gamma_0$ and $I_\textrm{sat}$ are taken as
fit parameters. Due to uncertainty in the power density
measurement and inhomogeneous intensity profile  we have to use
the second fit parameter for $I_\textrm{sat}$. Line widths
extrapolated to zero intensity are {
$\gamma_\textrm{410\,nm}(0)=10.5(2)$\,MHz} for the transition
$F_g=4\leftrightarrow F_e=5$ at 410.6\,nm and
{$\gamma_\textrm{420\,nm}(0)=3.8(1)$\,MHz} for the corresponding
hyperfine transition at 420.4\,nm. { The inhomogeneous radial
intensity distribution should modify the fit function in a complex
way. But substitution of the fit function (\ref{eq1}) by a linear
regression or by a function
$\gamma(I)=\gamma_0\sqrt{1+I/I_\textrm{sat}}$ changes extrapolated
values  for less than 0.2\,MHz.  We use this value as an
uncertainty resulting from our unprecise knowledge of the fit
function. } { Derivation of true saturation intensity and
$\gamma_0$  from the second fit parameter is hindered by the
complex radial intensity distribution in the laser beams. }

To evaluate the natural line widths of the transitions we correct
the obtained results for systematic broadenings. The main
contribution results from the laser line width which we evaluate
as 0.2(2)MHz according to the manufacturers specification (we
doubled the line width of the Ti:Sa laser assuming that the main
noise contribution results from slow long-term correlated acoustic
vibrations \cite{Rytov}). The next important broadening mechanism
is the time-of-flight broadening which we evaluate as 0.1(1)\,MHz.
The geometrical broadening resulting from the finite angle between
the beams and pressure shift can be conservatively estimated as
10(10)\,kHz and 50(50)\,kHz respectively.  {The Zeeman splitting
of $\pi$-components in a residual magnetic field is small due to
very similar magnetic $g$-factors of the lower and the upper
levels. The splitting of $\sigma$-components is about 1.5\,MHz/G.
For magnetic fields lower than 1\,G the Zeeman splittings
$\Delta_{Z}$ is relatively small ($\Delta_Z\ll\gamma_0$) and can
be considered similar to inhomogeneous spectral line broadening.
The following expression for the resulting line profile
$\gamma_\textrm{tot}$ is valid
$\gamma_\textrm{tot}\simeq\sqrt{\gamma_0^2+\Delta_{Z}^2}$ which is
similar to the case of Voigt function, when the contribution of
the Doppler width is much less than of the Lorentzian one. Thus
the Zeeman broadening of the transitions can be evaluated as
0.1(2)~MHz}. Taking all these contributions into account we
finally get the natural line width of {$10.0(4)$\,MHz} for the
transition
$4\textrm{f}^{13}6\textrm{s}^2(^2\textrm{F}_0)(F=4)\leftrightarrow4\textrm{f}^{12}(^3\textrm{H}_5)5\textrm{d}_{3/2}6\textrm{s}^2
(F=5)$ at 410.6\,nm  and {3.3(4)\,MHz} for the transition
$4\textrm{f}^{13}6\textrm{s}^2(^2\textrm{F}_0)(F=4)\leftrightarrow4\textrm{f}^{12}(^3\textrm{F}_4)5\textrm{d}_{5/2}6\textrm{s}^2(F=5)$
at 420.4\,nm. Our results are in a good agreement with transition
rates given in \cite{Wickliffe,NISTdatabase} and are summarized in
the next Section in table~\ref{tab2}.

\section{Optical leaks}

The candidates for cooling transitions at $\lambda=410.6$\,nm and
$\lambda=420.4$\,nm  are not perfectly closed. The excited levels
$4\textrm{f}^{12}(^3\textrm{H}_5)5\textrm{d}_{3/2}6\textrm{s}^2(J=9/2)$
and
$4\textrm{f}^{12}(^3\textrm{F}_4)5\textrm{d}_{5/2}6\textrm{s}^2
(J=9/2)$ can decay via electric dipole transitions to 6 and 4
neighboring opposite-parity levels respectively as shown in
Fig.\,\ref{fig1}. A crucial parameter for a cooling transition is
a branching ratio
\begin{equation}\label{eq2}
k=\frac{\sum A_i}{A_1+ \sum A_i},
\end{equation}
where $A_1$ is the decay rate to the ground state and $A_i$
($i=2,\ldots,5$ for the 420.4\,nm transition or $i=2,\ldots,7$ for
410.6\,nm one) are decay rates via other dipole transitions. In
the expression (\ref{eq2}) only the decay rate to the ground state
can be taken from \cite{Wickliffe,NISTdatabase} while all other
values remain unknown.

We evaluated the decay rates using the relativistic Hartree-Fock
code of Cowan \cite{Cowan}. The excited states have mixed
electronic configurations (see \cite{NISTdatabase}), and we took
into account a few leading configurations for each level. Since it
is difficult to achieve correct energies for all levels
simultaneously, we performed two sets of calculations. In the fist
set, the experimental energy of
$4\textrm{f}^{12}(^3\textrm{H}_5)5\textrm{d}_{3/2}6\textrm{s}^2(J=9/2)$
level ($E=24\,349$\,cm$^{-1}$) was taken as a reference for
calculations after which the required decay rates were calculated.
Similar evaluations where made for the
$4\textrm{f}^{12}(^3\textrm{F}_4)5\textrm{d}_{5/2}6\textrm{s}^2
(J=9/2)$ level  ($E=23\,782$\,cm$^{-1}$). Results of calculations
are summarized in table \ref{tab2}.

\begin{table}[h!]
\caption{Decay rates of  two excited levels
$4\textrm{f}^{12}(^3\textrm{H}_5)5\textrm{d}_{3/2}6\textrm{s}^2$
($E=24\,349\,\textrm{cm}^{-1}$) and
$4\textrm{f}^{12}(^3\textrm{F}_4)5\textrm{d}_{5/2}6\textrm{s}^2$
($E=23\,782\,\textrm{cm}^{-1}$) in atomic thulium. Energies
$E^\textrm{COWAN}_{g,e}$ and corresponding electric dipole
transition rates  $A^\textrm{COWAN}$ are calculated using
relativistic code of Cowan  \cite{Cowan}. For comparison, we give
the energies $E^\textrm{W}_{g,e}$ and rates $A^\textrm{W}$
measured in \cite{Wickliffe} and experimental results of this work
$A$ (the right column). }\label{tab2}
\begin{tabular}{ccccccccc}
 \hline \noalign{\smallskip}
  $E^\textrm{COWAN}_g$, & $E^\textrm{W}_g$,& $J_g$  & $E^\textrm{COWAN}_e$, & $E^\textrm{W}_e$,& $J_e$ & $A^\textrm{COWAN}$, &$A^\textrm{W}$, & $A$ (this work),\\\rule{0pt}{3ex}
 $10^3$\,cm$^{-1}$ & $10^3$\,cm$^{-1}$&  &$10^3$\,cm$^{-1}$ & $10^3$\,cm$^{-1}$&&s$^{-1}$&s$^{-1}$ &s$^{-1}$\\
   \noalign{\smallskip}\hline \noalign{\smallskip}
    0.000  & 0.000  & 3.5 & 24.341 & 24.349 & 4.5 & 2.13$\cdot10^8$&6.36(30)$\cdot10^7$&{$6.3(3)\cdot10^7$}\\
    22.166 & 22.420 & 4.5 & 24.341 & 24.349 & 4.5 & 4.44$\cdot10^2$&&\\
    22.243 & 22.468 & 5.5 & 24.341 & 24.349 & 4.5 & 1.75$\cdot10^2$&&\\
    22.417 & 22.560 & 5.5 & 24.341 & 24.349 & 4.5 & 1.82$\cdot10^2$&&\\
    22.905 & 23.335 & 3.5 & 24.341 & 24.349 & 4.5 & 1.38$\cdot10^2$&&\\
    23.622 & 23.941 & 4.5 & 24.341 & 24.349 & 4.5 & 1.54$\cdot10^1$&&\\
    23.893 & 23.873 & 3.5 & 24.341 & 24.349 & 4.5 & 2.95$\cdot10^0$&&\\
  \noalign{\smallskip}\hline \noalign{\smallskip}
    0.000  & 0.000  & 3.5 & 23.797 & 23.782 & 4.5 & 2.27$\cdot10^7$&2.43(12)$\cdot10^7$&{$2.1(3)\cdot10^7$}\\
    22.166 & 22.420 & 4.5 & 23.797 & 23.782 & 4.5 & 1.88$\cdot10^1$&&\\
    22.243 & 22.468 & 5.5 & 23.797 & 23.782 & 4.5 & 1.81$\cdot10^2$&&\\
    22.417 & 22.560 & 5.5 & 23.797 & 23.782 & 4.5 & 8.64$\cdot10^2$&&\\
    22.905 & 23.335 & 3.5 & 23.797 & 23.782 & 4.5 & 1.13$\cdot10^0$&&\\
   \noalign{\smallskip}\hline
\end{tabular}
\end{table}

Calculated energies are in a good agreement with experimental
data. Comparing calculated rates of strong transitions to the
ground state with ones from \cite{Wickliffe,NISTdatabase} we
observe a significant discrepancy for the transition at 410.6\,nm
(the first row in table~\ref{tab2}). We will take it as a scale
for the accuracy of our evaluation. Note, that the transition
$4\textrm{f}^{12}(^3\textrm{F}_4)5\textrm{d}_{5/2}6\textrm{s}^2\,(J=9/2)\rightarrow
4\textrm{f}^{13}(^2\textrm{F}_{7/2})5\textrm{d}6\textrm{s}\,(J=11/2)$
demonstrates the exceptionally high rate of about
$10^3$\,s$^{-1}$.

Using the results of table\,\ref{tab2} we evaluate the branching
ratios according to equation (\ref{eq2}):
{
\begin{eqnarray}
k_{410\,\textrm{nm}}=1_{-0.5}^{+1}\,\cdot10^{-5}\,,\label{eq3} \\
k_{420\,\textrm{nm}}=5_{-2.5}^{+5}\cdot10^{-5}\,,\label{eq4}
\end{eqnarray}}
where the uncertainties are evaluated according to the discrepancy
between calculated and experimental values. To get these values we
take experimental rates for $A_1$, while other rates $A_i$ are
taken from calculations. The given uncertainty is a realistic
estimation for the accuracy of our calculations.
 It would be
necessary to mention that not all of the atoms which have decayed
from the excited odd levels to the neighboring even levels are
taken away from a laser cooling cycle. Part of them can return
back to the ground state by cascade transitions while another part
sticks in metastable levels. Calculations of cascaded transitions
are a complex task, and it is reasonable to use the approach given
in the paper \cite{McClelland}, where these excited levels are
considered as a ``reservoir'' slowly feeding the ground state. We
use evaluations (\ref{eq3}), (\ref{eq4}) as upper limits for the
optical leak rates.

\section{Cooling transition}

We have analyzed the possibility  to cool atomic thulium from the
thermal beam at a temperature of 1100\,K using blue resonance
light in a Zeeman cooler.  Because of the relatively low
transition rate and the significant rate of optical leaks, the
transition at 420.4\,nm looks unfavorable for laser cooling.
Indeed, to completely decelerate a thulium atom with the initial
velocity of 200\,m/s it is necessary to have about {35\,000}
scattering events. For the transition at 420.4\,nm {97\%} of atoms
will be lost during deceleration if we take
$k_\textrm{420~nm}=10^{-4}$ (\ref{eq4}). On the other hand,
transition at 410.6\,nm is more suitable for cooling: for the
worst case estimation only {50\%} of the atoms will be lost. We
have numerically modelled the Zeeman slower of 40\,cm long
\cite{Barrett} and derived, that about 7\% of initial number of
atoms from the thermal beam can be decelerated to velocities of
20\,m/s with all optical leaks (table\,\ref{tab2}) taken into
account. We expect that using the transition at 410.6\,nm one can
cool and trap in a MOT up to $10^6$ thulium atoms with a loading
rate of about 1\,s for realistic oven and MOT parameters. The
repumping field, if necessary, can be produced from the cooling
field using an acousto-optical modulator operating at about
360\,MHz (see table \ref{tab1}).

The Doppler limit of this transition $T_\textrm{dop}=\hbar A/2k_B$
(here $k_B$ is the Boltzman constant) corresponds to 230\,$\mu$K
which is too high for successive loading in an optical dipole trap
or for making experiments in a ballistic flight. Further cooling
can be achieved by e.g. Sisyphus cooling or switching to a green
MOT at 530.7\,nm with $T_\textrm{dop}=9\,\mu$K.

\section{Clock transition}

As indicated in the Introduction, the transition between the
fine-structure levels of the ground state
$(J_g=7/2)\rightarrow(J'_g=5/2)$ at 1.14\,$\mu$m  can be
considered as a candidate for a clock transition due to its low
sensitivity to collisions and low differential polarizability of
the two stayes. In the work of Aleksandrov {\it et al.}
\cite{Aleksandrov} this transition has been observed in the
absorption spectrum of thulium vapors  using a high-resolution
spectrometer. Even after adding up to 50 bar of helium to a
thulium cell, the authors could not detect any spectral broadening
of this transition. They have concluded, that the pressure
broadening is less then 20\,MHz/bar which is 500 times less then
typical broadening of resonance s\,--\,p or f\,--\,d transitions
($\sim$10\,GHz/bar).

Using the Cowan code we have evaluated the transition rate at
$1.14\,\mu$m. The magnetic-dipole transition has a rate of
$A=5.9\,\textrm{s}^{-1}$ while the electric quadrupole transition
rate is negligible. The same evaluation has been done with the
help of the Flexinle Atomic Code of Gu Ming Feng \cite{FAC} which
results in $A=7.7\,\textrm{s}^{-1}$. The results agree with each
other and we give the final estimation of
$A=7(2)\,\textrm{s}^{-1}$ for this transition rate which
corresponds to the spectral line width of 1.1(3)\,Hz and the
transition $Q$-factor of $2.4(7)\cdot10^{14}$.
 Detection of the narrow unperturbed
resonance in a cold atomic cloud produced in MOT should increase
the signal to noise ratio and stability of a frequency reference
\cite{Katori}. The transition can be excited by a frequency
stabilized ytterbium fiber laser (see e.g. \cite{Kurkov} and
references therein) or, probably, by a stabilized
chomium:forsterite laser (Cr:Mg$_2$SiO$_4$) \cite{Sennaroglu}.

\section{Conclusions}

In this work we have analyzed the possibility to cool atomic
thulium using strong blue transitions in the spectral range
410\,--\,420\,nm. The hyperfine structure of two candidates for
cooling transitions at 410.6\,nm and 420.4\,nm and of two other
transitions at 409.5\,nm and 418.9\,nm is determined by means of
saturation absorption spectroscopy. From analysis of the spectral
line widths we derived the corresponding transition rates with an
accuracy better then 10\%. We have evaluated the role of optical
leaks by calculation of decay rates from the excited levels
$4\textrm{f}^{12}(^3\textrm{H}_5)5\textrm{d}_{3/2}6\textrm{s}^2$
($E=24\,349\,\textrm{cm}^{-1}$) and
$4\textrm{f}^{12}(^3\textrm{F}_4)5\textrm{d}_{5/2}6\textrm{s}^2$
($E=23\,782\,\textrm{cm}^{-1}$) using the code of Cowan
\cite{Cowan}.

We conclude, that the transition
$4\textrm{f}^{13}6\textrm{s}^2(^2\textrm{F}_0)(F_g=4)
\leftrightarrow4\textrm{f}^{12}(^3\textrm{F}_4)5\textrm{d}_{5/2}6\textrm{s}^2
(F_e=5)$ at 410.6\,nm can be used for effective laser cooling of
thulium from a hot atomic beam. Evaluations show, that
deceleration of atomic beam in a 40-cm Zeeman slower will allow
one to decelerate about 5\% of atoms from the thermal distribution
and to trap up to $10^6$ atoms in a magneto-optical trap. Further
cooling can be achieved in a green MOT operating at
$\lambda=530.7$\,nm.

We have evaluated the rate of  the electric-dipole forbidden
transition at 1.14\,$\mu$m between the fine structure levels of
the ground state $(J_g=7/2)\rightarrow(J'_g=5/2)$. A calculated
rate of the magnetic-dipole transition is
$A=7(2)\,\textrm{s}^{-1}$ which corresponds to the quality factor
of $2\cdot10^{14}$. { This shielded transition can be considered
as one of the candidates for applications in optical atomic
clocks.}

\section*{Acknowledgments}

The work is partly supported by the Alexander von Humboldt
Foundation, Russian Science Support Foundation and RFBR Grants
\#05-02-16801, \#08-02-00667, \#05-02-00443.

\end{document}